\documentclass[11pt,english]{llncs}
\usepackage[T1]{fontenc}
\usepackage[latin1]{inputenc}
\usepackage{array}
\usepackage{amssymb}

\makeatletter

\providecommand{\tabularnewline}{\\}

\usepackage{stmaryrd}
\newcommand{\llb}{\llbracket}
\newcommand{\rrb}{\rrbracket}

\usepackage{babel}
\makeatother
\begin{document}

\vfill{}
\title{Yet Another Normalisation Proof for Martin-L\"{o}f's Logical Framework}
\vfill{}

\subtitle{--- Terms with correct arities are \\
strongly normalising}

\author{Yong Luo}

\institute{Computing Laboratory, University of Kent, Canterbury, UK\\
Email: Y.Luo@kent.ac.uk}

\maketitle
\begin{abstract}
In this paper, we prove the strong normalisation for Martin-L\"{o}f's
Logical Framework, and suggest that {}``correct arity'', a condition
weaker than well-typedness, will also guarantee the strong normalisation.
\end{abstract}

\section{Introduction}

The normalisation proofs for dependently typed systems are known to
be notoriously difficult. For example, if we have a task to prove
strong normalisation for Martin-L\"{o}f's Logical Framework (MLF)
(in the Appendix), and if we use typed operational semantics as in
\cite{healf:thesis}, the proof would be more than one hundred pages
long. When a proof is long and complicated, it is likely found to
contain mistakes and bugs \cite{coq:thesis,CoqGal:SN,alti:types94}.
This paper presents an elegant and comprehensible proof of strong
normalisation for MLF.

We often associate well-typedness with strong normalisation in type
systems. But this paper suggests that well-typedness may have little
to do with strong normalisation in essence, and proves that terms
with correct arities are strongly normalising. The condition of {}``correct
arity'' is weaker than that of well-typedness (\emph{i.e.} well-typed
terms have correct arities). This paper will also demonstrate the
difference between types and arities when we extend MLF with some
inductive data types and their computation rules. New reduction rules
will not increase the set of terms with correct arities, but they
will usually increase the set of well-typed terms. One of the reasons
is that there are reductions inside types (\emph{i.e.} one type can
be reduced to another type) in a dependently typed system but there
is no reduction for arities.

Our goal is to prove the strong normalisation w.r.t. $\beta$ and
$\eta$-reduction. But it is very difficult to prove it directly.
An important technique in the paper is that, we extend the definition
of terms and kinds, and introduce a new reduction rule $\beta_{2}$
for kinds. Then, we prove a stronger and more general property, that
is, strong normalisation w.r.t. $\beta$, $\eta$ and $\beta_{2}$-reduction.
In this way, the proof becomes easier although the property is stronger.
Without the $\beta_{2}$-reduction, the proof of soundness in Section
\ref{sec:Normalisation-proof} is impossible to go through.

In Section \ref{sec:Basic-definitions}, we give some basic definitions
that are used throughout the paper. In Section \ref{sec:Inference-rules},
the inference rules of arities are formally presented. In Section
\ref{sec:Normalisation-proof}, we give more definitions such as saturated
sets, and prove the strong normalisation for the arity system. In
Section \ref{sec:Computation-rules}, the computation rules for the
type of dependent pairs and finite types and simple computation rules
for universes are introduced. The strong normalisation for a dependently
typed system is proved by the commutation property between these rules
and $\beta$-reduction. The conclusions and future work are discussed
in the last section.

\subsubsection*{Related work}

Logical frameworks arise because one wants to create a single framework,
which is a kind of meta-logic or universal logic. The Edinburgh Logical
Framework \cite{LF:87,ELF:92} presents logics by a \emph{judgements-as-types}
principle, which can be regarded as the meta-theoretical analogue
of the well-known \emph{propositions-as-types} principles \cite{curryfeys:book,deBruijn:Automath80,How:69}.
Martin-L\"{o}f's logical framework \cite{ML:book84,NPS:book} has
been developed by Martin-L\"{o}f to present his intensional type
theory. In UTT \cite{luo:book94}, Luo proposed a typed version of
Martin-L\"{o}f's logical framework, in which untyped functional operations
of the form $(x)k$ are replaced by typed $[x:K]k$.

There are many normalisation proofs for simply typed systems and dependently
typed systems in literature \cite{barendregt:PTShandbook92,luo:thesis,thorsten:thesis}
\cite{paul_werner,healf:thesis} \cite{geuvers:thesis,Werner:SN}.
The techniques employed in this paper such as the interpretation of
arities and saturated sets are inspired by and closely related to
the proof for simply typed calculus in \cite{barendregt:PTShandbook92}.
The concept of arity is well-known in mathematics and it is often
defined as the maximum number of arguments that a function can have.
But in this paper, the definition of arity and the concept of {}``correct
arity'' are different. The complexity of the normalisation proof
for MLF is dramatically decreased because of this concept and other
techniques such as a new case of kinds and the corresponding $\beta_{2}$-reduction.
The commutation property was also studied in literature such as \cite{barendregt:lambda-book,Cosmo}.
The properties of Church-Rosser and strong normalisation for finite
types in simply typed systems are also studied in \cite{Sergei:Iso}.

\section{\label{sec:Basic-definitions}Basic definitions}

In this section, we give some basic definitions that will be used
later, and give the redice and the corresponding reduction rules.

\begin{definition}
\textbf{(Terms and Kinds)}
\begin{itemize}
\item Terms

\begin{enumerate}
\item a variable is a term,
\item $\lambda x:K.M$ is a term if $x$ is a variable, $K$ is a kind and
$M$ is a term,
\item $MN$ is a term if $M$ and $N$ are terms.
\end{enumerate}
\item Kinds

\begin{enumerate}
\item $Type$ is a kind,
\item $El(M)$ is a kind if $M$ is a term,
\item $(x:K_{1})K_{2}$ is a kind if $K_{1}$ and $K_{2}$ are kinds,
\item $KN$ is a kind if $K$ is a kind and $N$ is a term.
\end{enumerate}
\end{itemize}
\end{definition}
\begin{remark}
Terms and kinds are mutually and recursively defined. This definition
allows more terms and kinds than that of MLF since the forth case
for the definition of kinds is not included in MLF (see Appendix for
details).
\end{remark}
\begin{description}
\item [Notation:]Following the tradition, $\Lambda$ denotes the set of
all terms and $\Pi$ the set of all kinds. We sometimes write $f(a)$
for $fa$, $f(a,b)$ for $(fa)b$ and so on. $[N/x]M$ stands for
the expression obtained from $M$ by substituting $N$ for the free
occurrences of variable $x$ in $M$. $FV(M)$ is the set of free
variables in $M$.
\end{description}

\subsection*{Redice and reduction rules}

There are three different forms of redice: $(\lambda x:K.M)N$, $((x:K_{1})K_{2})N$
and $\lambda x:K.Mx$ when $x\not\in FV(M)$. The reduction rules
for these redice are the following.

\[
(\lambda x:K.M)N\longrightarrow_{\beta}[N/x]M\]
\[
((x:K_{1})K_{2})N\longrightarrow_{\beta_{2}}[N/x]K_{2}\]
\[
\lambda x:K.Mx\longrightarrow_{\eta}M\,\,\,\,\,\,\,\,\,\, x\not\in FV(M)\]

\begin{remark}
The second rule $\longrightarrow_{\beta_{2}}$ is new and is not included
in MLF. This rule will make the soundness proof go through easily
although the property is stronger and more general.
\end{remark}
\begin{description}
\item [Notation:]$\longrightarrow_{R}$ represents one-step $R$-reduction,
precisely, $M\longrightarrow_{R}N$ if a sub-term $P$ of $M$ is
a $R$-redex and $N$ is obtained by replacing $P$ by the result
after applying the reduction rule $R$. $M\twoheadrightarrow_{R}N$
means there is $0$ or more but finite steps of $R$-reduction from
$M$ to $N$. $M\twoheadrightarrow_{R}^{+}N$ means there is at least
one but finite steps of $R$-reduction from $M$ to $N$.
\end{description}
\begin{definition}
\textbf{(Arities)}
\begin{itemize}
\item $Zero$ is an arity,
\item $(a_{1},a_{2})$ is an arity if $a_{1}$ and $a_{2}$ are arities.
\end{itemize}
\end{definition}
\begin{description}
\item [Notation:]$\Omega$ denotes the set of all arities.
\end{description}

\section{\label{sec:Inference-rules}Inference rules}

In this section, we formally present the inference rules of arities.

The judgement form will be the following form,\[
A\vdash M:a\]
where $A\equiv<x_{1}:a_{1},...,x_{n}:a_{n}>$ is a finite sequence
of $x_{i}:a_{i}$, $x_{i}$ is a variable and $a_{i}$ is an arity;
$M$ is a term or kind; and $a$ is an arity. We shall read this judgement
like {}``under the context $A$, the term or kind $M$ has arity
$a$''.

\begin{description}
\item [Notation]For a context $A\equiv x_{1}:a_{1},...,x_{n}:a_{n}$, $FV(A)$
represents the set $\{ x_{1},...,x_{n}\}$.
\end{description}
All of the inference rules of arities are in Figure \ref{fig:arity}.

\begin{figure}
\begin{tabular}{|p{12cm}|}
\hline 
~

\textbf{Contexts:}

\[
\frac{}{<>\,\,\, valid}\,\,\,\,\,\,\,\,\,\,\,\,\frac{A\,\,\, valid\,\,\,\,\,\,\, x\not\in FV(A)\,\,\,\,\,\, a\in\Omega}{A,x:a}\]

~

\textbf{Inference rules for kinds:}

\[
\frac{A\,\,\, valid}{A\vdash Type:Zero}\,\,\,\,\,\,\,\,\,\,\,\,\,\,\frac{A\vdash M:Zero}{A\vdash El(M):Zero}\]

\[
\frac{A\vdash K_{1}:a_{1}\,\,\,\,\,\, A,x:a_{1}\vdash K_{2}:a_{2}}{A\vdash(x:K_{1})K_{2}:(a_{1},a_{2})}\,\,\,\,\,\,\,\,\,\,\,\,\,\,\,\frac{A\vdash K:(a_{1},a_{2})\,\,\,\,\,\, A\vdash N:a_{1}}{A\vdash KN:a_{2}}\]

~

\textbf{Inference rules for terms:}

\textbf{\[
\frac{A,x:a,A'\,\,\, valid}{A,x:a,A'\vdash x:a}\]
}

\[
\frac{A\vdash K:a_{1}\,\,\,\,\,\, A,x:a_{1}\vdash M:a_{2}}{A\vdash\lambda x:K.M:(a_{1},a_{2})}\,\,\,\,\,\,\,\,\,\,\,\,\,\,\,\,\frac{A\vdash M:(a_{1},a_{2})\,\,\,\,\,\, A\vdash N:a_{1}}{A\vdash MN:a_{2}}\]

~\tabularnewline
\hline
\end{tabular}

\caption{\label{fig:arity}Inference rules of arities}
\end{figure}

\begin{definition}
We say that a term or kind $M$ has \textbf{a correct arity} if $A\vdash M:a$
is derivable for some $A$ and $a$.
\end{definition}
\begin{remark}
We have the following remarks:
\begin{itemize}
\item A well-typed term has a correct arity (a proof will be given later),
but a term which has a correct arity is not necessarily well-typed.
For instance, under the context\[
A:Type,B:Type,C:Type,f:(x:A)C,b:B\]
the term $f(b)$ is not well-typed, but it has a correct arity $Zero$
under the following context\[
A:Zero,B:Zero,C:Zero,f:(Zero,Zero),b:Zero\]
 Another example with dependent type is that, under the context\[
\begin{array}{c}
A:Type,B:(x:A)Type,f:(x:A)(y:B(x))Type,\\
x_{1}:A,x_{2}:A,b:B(x_{2})\end{array}\]
the term $f(x_{1},b)$ is not well-typed, but it has a correct arity
$Zero$ in the following context\[
\begin{array}{c}
A:Zero,B:(Zero,Zero),f:(Zero,(Zero,Zero)),\\
x_{1}:Zero,x_{2}:Zero,b:Zero\end{array}\]

\item For any judgement $A\vdash M:a$, $M$ must be either a kind or a
term. A derivation such as $\frac{A\vdash Type:Zero}{A\vdash El(Type):Zero}$
is not possible, because $El(Type)$ is neither a term nor a kind.
\end{itemize}
\end{remark}
\begin{lemma}
\label{lem:selfapp}If both $A\vdash M:a$ and $A\vdash M:b$ are
derivable then $a$ and $b$ are syntactically the same ($a\equiv b$).
And $A\vdash MM:a$ is not derivable for any $A$, $M$ and $a$.
\end{lemma}
\begin{proof}
By induction on the derivations of $A\vdash M:a$ and $A\vdash M:b$.
\end{proof}
\begin{remark}
One may recall that the non-terminating example $\omega\omega$ where
$\omega\equiv\lambda x.xx$. It is impossible that $\omega$ is well-typed
in a simply typed calculus \cite{barendregt:PTShandbook92}. By Lemma
\ref{lem:selfapp}, it is also impossible to have a correct arity
for $\omega$.
\end{remark}

\section{\label{sec:Normalisation-proof}Normalisation proof}

In this section, we give more definitions such as saturated sets to
prove the strong normalisation for the arity system.

\begin{definition}
\textbf{(Interpretation of arities)}
\begin{itemize}
\item $SN^{\Lambda}=_{df}\{ M\in\Lambda\,\,|\,\, M\, is\, strongly\, normalising\}$.
\item $SN^{\Pi}=_{df}\{ M\in\Pi\,\,|\,\, M\, is\, strongly\, normalising\}$.
\item $\llb Zero\rrb^{\Lambda}=_{df}SN^{\Lambda}$.
\item $\llb Zero\rrb^{\Pi}=_{df}SN^{\Pi}$.
\item $\llb(a_{1},a_{2})\rrb^{\Lambda}=_{df}\{ M\in\Lambda\,\,|\,\,\forall N\in\llb a_{1}\rrb^{\Lambda},\,\, MN\in\llb a_{2}\rrb^{\Lambda}\}$.
\item $\llb(a_{1},a_{2})\rrb^{\Pi}=_{df}\{ K\in\Pi\,\,|\,\,\forall N\in\llb a_{1}\rrb^{\Lambda},\,\, KN\in\llb a_{2}\rrb^{\Pi}\}$.
\end{itemize}
\end{definition}
\begin{remark}
$\llb a\rrb^{\Lambda}$ is a set of terms, while $\llb a\rrb^{\Pi}$
is a set of kinds for any arity $a$.
\end{remark}
\begin{description}
\item [Notations:]We shall write $\overline{R}$ for $R_{1},R_{2},...,R_{n}$
for some $n\geq0$, and $M\overline{R}$ for $(...((MR_{1})R_{2})...R_{n})$.
\end{description}
\begin{definition}
\textbf{(Saturated sets)}
\begin{itemize}
\item A subset $X\subseteq SN^{\Lambda}$ is called saturated if

\begin{enumerate}
\item $\forall\overline{R}\in SN^{\Lambda}$, $x\overline{R}\in X$ where
$x$ is any term variable,
\item $\forall\overline{R}\in SN^{\Lambda}$, $\forall Q\in SN^{\Lambda}$
and $\forall K\in SN^{\Pi}$,\[
([Q/x]P)\overline{R}\in X\Longrightarrow(\lambda x:K.P)Q\overline{R}\in X\]

\end{enumerate}
\item A subset $Y\subseteq SN^{\Pi}$ is called saturated if $\forall\overline{R}\in SN^{\Lambda}$,
$\forall N\in SN^{\Lambda}$ and $\forall K_{1}\in SN^{\Pi}$,\[
([N/x]K_{2})\overline{R}\in Y\Longrightarrow((x:K_{1})K_{2})N\overline{R}\in Y\]

\item $SAT^{\Lambda}=_{df}\{ X\subseteq SN^{\Lambda}\,\,|\,\, X\, is\, saturated\}$
\item $SAT^{\Pi}=_{df}\{ Y\subseteq SN^{\Pi}\,\,|\,\, Y\, is\, saturated\}$
\end{itemize}
\end{definition}
\begin{lemma}
\textbf{(Arities and saturated sets)}
\begin{itemize}
\item $SN^{\Lambda}\in SAT^{\Lambda}$ and $SN^{\Pi}\in SAT^{\Pi}$.
\item $a\in\Omega\Longrightarrow\llb a\rrb^{\Lambda}\in SAT^{\Lambda}$
and $\llb a\rrb^{\Pi}\in SAT^{\Pi}$.
\end{itemize}
\end{lemma}
\begin{proof}
By the definition of saturated sets and by induction on arities.
\begin{itemize}
\item Let's prove $SN^{\Lambda}\in SAT^{\Lambda}$ first. We have $SN^{\Lambda}\subseteq SN^{\Lambda}$
and $x\overline{R}\in SN^{\Lambda}$ if $\overline{R}\in SN^{\Lambda}$.
Now we need to prove for $Q,\overline{R}\in SN^{\Lambda}$ and $K\in SN^{\Pi}$,
\[
([Q/x]P)\overline{R}\in SN^{\Lambda}\Longrightarrow(\lambda x:K.P)Q\overline{R}\in SN^{\Lambda}\]
Since $([Q/x]P)\overline{R}\in SN^{\Lambda}$, we have $P\in SN^{\Lambda}$
and after any finitely many steps reducing inside $P$, $Q$ and $\overline{R}$,
$([Q'/x]P')\overline{R'}\in SN^{\Lambda}$ with $P\twoheadrightarrow_{\beta\eta}P'$
, $Q\twoheadrightarrow_{\beta\eta}Q'$ and $\overline{R}\twoheadrightarrow_{\beta\eta}\overline{R'}$.\\
From $(\lambda x:K.P)Q\overline{R}$, after any finitely many steps
reducing inside $P$, $Q$, $\overline{R}$ and $K$, and we get $(\lambda x:K'.P')Q'\overline{R'}$.
From here, we may have two choices.

\begin{itemize}
\item $(\lambda x:K'.P')Q'\overline{R'}\longrightarrow_{\beta}([Q'/x]P')\overline{R'}$
\item $P'\equiv Fx$ and $x\not\in FV(F)$ and\[
(\lambda x:K'.P')Q'\overline{R'}\longrightarrow_{\eta}FQ'\overline{R'}\equiv([Q'/x]P')\overline{R'}\]

\end{itemize}
For both cases, because $([Q'/x]P')\overline{R'}\in SN^{\Lambda}$,
we have $(\lambda x:K.P)Q\overline{R}\in SN^{\Lambda}$.

\item The proof of $SN^{\Pi}\in SAT^{\Pi}$ is similar to that of $SN^{\Lambda}\in SAT^{\Lambda}$.
\item Now, let's prove $\llb a\rrb^{\Lambda}\in SAT^{\Lambda}$ by induction
on $a$. The base case (\emph{i.e.} $\llb Zero\rrb^{\Lambda}=SN^{\Lambda}\in SAT^{\Lambda}$)
has been proved. So we only need to prove $\llb(a_{1},a_{2})\rrb^{\Lambda}\in SAT^{\Lambda}$.
By induction hypothesis, we have $\llb a_{1}\rrb^{\Lambda}\in SAT^{\Lambda}$
and $\llb a_{2}\rrb^{\Lambda}\in SAT^{\Lambda}$.\\
Then we have $x\in\llb a_{1}\rrb^{\Lambda}$ for all variable $x$.
Therefore \begin{eqnarray*}
F\in\llb(a_{1},a_{2})\rrb^{\Lambda} & \Longrightarrow & Fx\in\llb a_{2}\rrb^{\Lambda}\\
 & \Longrightarrow & Fx\in SN^{\Lambda}\\
 & \Longrightarrow & F\in SN^{\Lambda}\end{eqnarray*}
So, we have $\llb(a_{1},a_{2})\rrb^{\Lambda}\subseteq SN^{\Lambda}$.\\
Now, we need to prove that for any variable $x$ and $\forall\overline{R}\in SN^{\Lambda}$,
we have $x\overline{R}\in\llb(a_{1},a_{2})\rrb^{\Lambda}$. This means\[
\forall N\in\llb a_{1}\rrb^{\Lambda}\,\,\,\,\, x\overline{R}N\in\llb a_{2}\rrb^{\Lambda}\]
which is true since $\llb a_{1}\rrb^{\Lambda}\subseteq SN^{\Lambda}$
and $\llb a_{2}\rrb^{\Lambda}\in SAT^{\Lambda}$.\\
Finally, we need to prove that for $\forall\overline{R}\in SN^{\Lambda}$,
$\forall Q\in SN^{\Lambda}$ and $\forall K\in SN^{\Pi}$,\[
([Q/x]P)\overline{R}\in\llb(a_{1},a_{2})\rrb^{\Lambda}\Longrightarrow(\lambda x:K.P)Q\overline{R}\in\llb(a_{1},a_{2})\rrb^{\Lambda}\]
Since $([Q/x]P)\overline{R}\in\llb(a_{1},a_{2})\rrb^{\Lambda}$, we
have $([Q/x]P)\overline{R}N\in\llb a_{2}\rrb^{\Lambda}$ for $\forall N\in\llb a_{1}\rrb^{\Lambda}$.
And since $\llb a_{1}\rrb^{\Lambda}\subseteq SN^{\Lambda}$ and $\llb a_{2}\rrb^{\Lambda}\in SAT^{\Lambda},$
we have $(\lambda x:K.P)Q\overline{R}N\in\llb a_{2}\rrb^{\Lambda}$
and hence\[
(\lambda x:K.P)Q\overline{R}\in\llb(a_{1},a_{2})\rrb^{\Lambda}\]

\item The proof of $\llb a\rrb^{\Pi}\in SAT^{\Pi}$ is similar to that of
$\llb a\rrb^{\Lambda}\in SAT^{\Lambda}$\qed 
\end{itemize}
\end{proof}
\begin{description}
\item [Notation:]We often use $SN$ for $SN^{\Lambda}\cup SN^{\Pi}$ and
$\llb a\rrb$ for $\llb a\rrb^{\Lambda}\cup\llb a\rrb^{\Pi}$.
\end{description}
\begin{definition}
\textbf{(Valuation)}
\begin{itemize}
\item A \emph{valuation} is a map $\rho:V\rightarrow\Lambda$, where $V$
is the set of all term variables.
\item Let $\rho$ be a valuation. Then\[
\llb M\rrb_{\rho}=_{df}[\rho(x_{1})/x_{1},...,\rho(x_{n})/x_{n}]M\]
where $x_{1},...,x_{n}$ are all of the free variable in $M$.
\item Let $\rho$ be a valuation. Then 

\begin{itemize}
\item $\rho$ satisfies $M:a$, notation $\rho\models M:a,$ if $\llb M\rrb_{\rho}\in\llb a\rrb$;
\item $\rho$ satisfies $A$, notation $\rho\models A$, if $\rho\models x:a$
for all $x:a\in A$;
\item $A$ satisfies $M:a$, notation $A\models M:a,$ if\[
\forall\rho\,\,(\rho\models A\Longrightarrow\rho\models M:a)\]

\end{itemize}
\end{itemize}
\end{definition}
\begin{remark}
For any valuation $\rho$, if $M$ is a term, $\llb M\rrb_{\rho}$
is also a term, and similarly, if $M$ is a kind, $\llb M\rrb_{\rho}$
is also a kind. If a valuation $\rho$ satisfies that $\rho(x)=x$
then $\llb M\rrb_{\rho}\equiv M$.
\end{remark}
\begin{lemma}
\textbf{\label{lem:Soundness}(Soundness) $A\vdash M:a\Longrightarrow A\models M:a$}
where $M$ is a term or kind.
\end{lemma}
\begin{proof}
By induction on the derivations of $A\vdash M:a$.
\begin{enumerate}
\item The last rule is\[
\frac{A\,\,\, valid}{A\vdash Type:Zero}\]
 Since $\llb Type\rrb_{\rho}=Type$ for any $\rho$ and $Type\in SN=\llb Zero\rrb$,
we have $\llb Type\rrb_{\rho}\in\llb Zero\rrb$.
\item The last rule is\[
\frac{A\vdash M:Zero}{A\vdash El(M):Zero}\]
Since $\llb El(M)\rrb_{\rho}=El(\llb M\rrb_{\rho})$ for any $\rho$
and $\llb M\rrb_{\rho}\in\llb Zero\rrb=SN$, we have $\llb El(M)\rrb_{\rho}\in SN=\llb Zero\rrb$.
\item The last rule is\[
\frac{A\vdash K_{1}:a_{1}\,\,\,\,\,\, A,x:a_{1}\vdash K_{2}:a_{2}}{A\vdash(x:K_{1})K_{2}:(a_{1},a_{2})}\]
We must show that\[
\forall\rho\,\,(\rho\models A\Longrightarrow\rho\models(x:K_{1})K_{2}:(a_{1},a_{2}))\]
That is, we must show that $\llb(x:K_{1})K_{2}\rrb_{\rho}\in\llb(a_{1},a_{2})\rrb^{\Pi}$.
By the definition of $\llb(a_{1},a_{2})\rrb^{\Pi}$, we must show
that, for all $N\in\llb a_{1}\rrb^{\Lambda}$,\[
\llb(x:K_{1})K_{2}\rrb_{\rho}N\in\llb a_{2}\rrb^{\Pi}\]
Note that \begin{eqnarray*}
\llb(x:K_{1})K_{2}\rrb_{\rho}N & \equiv & ((x:K_{1}')K_{2}')N\\
 & \rightarrow_{\beta_{2}} & [N/x]K_{2}'\\
 & \equiv & \llb K_{2}\rrb_{\rho\cup(N/x)}\end{eqnarray*}
where $K_{1}'\equiv\llb K_{1}\rrb_{\rho}\equiv[\rho(y_{i})/y_{i}...]K_{1}$
and $K_{2}'\equiv\llb K_{2}\rrb_{\rho}\equiv[\rho(y_{i})/y_{i}...]K_{2}$\\
Now, let's consider the induction hypothesis. Since $\rho\cup(N/x)\models A,x:a_{1}$,
we have $\llb K_{1}\rrb_{\rho}\in\llb a_{1}\rrb^{\Pi}$ and $\llb K_{2}\rrb_{\rho\cup(N/x)}\in\llb a_{2}\rrb^{\Pi}$.
So, we have $[N/x]K_{2}'\in\llb a_{2}\rrb^{\Pi}$, and because $\llb a_{2}\rrb^{\Pi}$
is saturated, we have $((x:K_{1}')K_{2}')N\in\llb a_{2}\rrb^{\Pi}$,
i.e. $\llb(x:K_{1})K_{2}\rrb_{\rho}N\in\llb a_{2}\rrb^{\Pi}$. Note
that, since $\llb a_{1}\rrb^{\Lambda}\subseteq SN^{\Lambda}$ and
$\llb a_{1}\rrb^{\Pi}\subseteq SN^{\Pi}$, we know that $N\in SN^{\Lambda}$
and $K_{1}'\in SN^{\Pi}$.
\item The last rule is\[
\frac{A\vdash K:(a_{1},a_{2})\,\,\,\,\,\, A\vdash N:a_{1}}{A\vdash KN:a_{2}}\]
We must show that\[
\forall\rho\,\,(\rho\models A\Longrightarrow\rho\models KN:a_{2})\]
By induction hypothesis, we have $\llb K\rrb_{\rho}\in\llb(a_{1},a_{2})\rrb^{\Pi}$
and $\llb N\rrb_{\rho}\in\llb a_{1}\rrb^{\Lambda}$.\\
By the definition of $\llb(a_{1},a_{2})\rrb^{\Pi}$, we have $\llb K\rrb_{\rho}\llb N\rrb_{\rho}\in\llb a_{2}\rrb^{\Pi}$,
\emph{i.e.} $\llb KN\rrb_{\rho}\in\llb a_{2}\rrb^{\Pi}$.
\item The last rule is\[
\frac{A,x:a,A'\,\,\, valid}{A,x:a,A'\vdash x:a}\]
Easy.
\item The last rule is\[
\frac{A\vdash K:a_{1}\,\,\,\,\,\, A,x:a_{1}\vdash M:a_{2}}{A\vdash\lambda x:K.M:(a_{1},a_{2})}\]
Similar to case 3.
\item The last rule is\[
\frac{A\vdash M:(a_{1},a_{2})\,\,\,\,\,\, A\vdash N:a_{1}}{A\vdash MN:a_{2}}\]
Similar to case 4. \qed
\end{enumerate}
\end{proof}
\begin{theorem}
\label{thm:SN}If $A\vdash M:a$, then $M$ is strongly normalising.
\end{theorem}
\begin{proof}
By Lemma \ref{lem:Soundness} and take the evaluation $\rho_{0}$
that satisfies $\rho_{0}(x)=x$. 

By Lemma \ref{lem:Soundness}, we have $A\models M:a$. So, by definition,
we have\[
\rho_{0}\models A\Longrightarrow\rho_{0}\models M:a\]
Suppose $A\equiv x_{1}:a_{1},...,x_{n}:a_{n}.$ Since $\llb a_{i}\rrb^{\Lambda}\in SAT^{\Lambda}$,
we have $x_{i}\in\llb a_{i}\rrb^{\Lambda}$. Hence $\rho_{0}\models A$.
So, we have $\rho_{0}\models M:a$ and hence $M=\llb M\rrb_{\rho_{0}}\in\llb a\rrb\subseteq SN$.\qed
\end{proof}

\subsection*{Translation from kinds to arities}

Now, we define a map to translate kinds to arities, and prove that
well-typed terms have correct arities.

\begin{definition}
A map $arity:\Pi\rightarrow\Omega$ is inductively defined as follows.
\begin{itemize}
\item $arity(Type)=Zero$,
\item $arity(El(A))=Zero$,
\item $arity((x:K_{1})K_{2})=(arity(K_{1}),arity(K_{2}))$.
\end{itemize}
\end{definition}
\begin{description}
\item [Notation:]Suppose a context $\Gamma\equiv x_{1}:K_{1},...,x_{n}:K_{n}$,
then $arity(\Gamma)\equiv x_{1}:arity(K_{1}),...,x_{n}:arity(K_{n})$.
\end{description}
\begin{theorem}
\textbf{\emph{\label{thm:type_arity}(Well-typed terms have correct
arities)}} If $\Gamma\vdash M:K$ is derivable in MLF, then $arity(\Gamma)\vdash M:arity(K)$
is derivable.
\end{theorem}
\begin{proof}
By induction on the derivations of $\Gamma\vdash M:K$ (see the inference
rules of MLF in Appendix).
\end{proof}
\begin{theorem}
If $\Gamma\vdash M:K$ is derivable in MLF, then $M$ is strongly
normalising.
\end{theorem}
\begin{proof}
By Theorem \ref{thm:SN} and Theorem \ref{thm:type_arity}.
\end{proof}

\section{\label{sec:Computation-rules}Computation rules}

In this section, we shall introduce computation rules for the type
of dependent pairs and finite types and simple computation rules for
universes. The strong normalisation is proved in a way that no one
has ever take before in dependently typed systems, to the author's
best knowledge. Recall that adding new computation (or reduction)
rules will not increase the set of terms with correct arities. The
basic strategy we adopt is to prove strong normalisation one reduction
rule after another. That is, if we have already proved strong normalisation
for a set of reduction rules, after adding one new reduction rule,
can we still prove strong normalisation? This strategy will not work
for dependently typed systems if we want to prove the statement that
{}``well-typed terms are strongly normalising'', because whenever
we add a single computation rule, the set of well-typed terms may
increase.

\subsection{The type of dependent pairs}

In MLF, the constants and computation rules for the type of dependent
pairs can be specified as follows:\begin{eqnarray*}
\Sigma & : & (A:Type)(B:(A)Type)Type\\
pair & : & (A:Type)(B:(A)Type)(a:A)(b:B(a))\Sigma(A,B)\\
\pi_{1} & : & (A:Type)(B:(A)Type)(z:\Sigma(A,B))A\\
\pi_{2} & : & (A:Type)(B:(A)Type)(z:\Sigma(A,B))B(\pi_{1}(A,B,z))\end{eqnarray*}
\begin{eqnarray*}
\pi_{1}(A,B,pair(A,B,a,b)) & = & a\,\,\,:\,\,\, A\\
\pi_{2}(A,B,pair(A,B,a,b)) & = & b\,\,\,:\,\,\, B(a)\end{eqnarray*}
In the arity system of the paper, we change the kinds to arities and
the constants and the reduction rules are introduced as the following:\begin{eqnarray*}
\Sigma & : & (Zero,((Zero,Zero),Zero))\\
pair & : & (Zero,((Zero,Zero),(Zero,(Zero,Zero))))\\
\pi_{1} & : & (Zero,((Zero,Zero),(Zero,Zero)))\\
\pi_{2} & : & (Zero,((Zero,Zero),(Zero,Zero)))\end{eqnarray*}
\begin{eqnarray*}
\pi_{1}(A,B,pair(A,B,a,b)) & \longrightarrow_{\pi_{1}} & a\,\,:\,\, Zero\\
\pi_{2}(A,B,pair(A,B,a,b)) & \longrightarrow_{\pi_{2}} & b\,\,:\,\, Zero\end{eqnarray*}

\subsection{Finite types}

In type systems, a finite type ${\cal T}$ can be represented by following
constants\begin{eqnarray*}
{\cal T} & : & Type\\
c_{1} & : & {\cal T}\\
. & . & .\\
c_{n} & : & {\cal T}\\
{\cal E_{{\cal T}}} & : & (P:({\cal T})Type)\\
 &  & (P(c_{1}))...(P(c_{n}))\\
 &  & (z:{\cal T})(P(z))\end{eqnarray*}
and the following computation rules\begin{eqnarray*}
{\cal E_{{\cal T}}}(P,p_{1},...,p_{n},c_{1}) & = & p_{1}\,\,:\,\, P(c_{1})\\
......\\
{\cal E_{{\cal T}}}(P,p_{1},...,p_{n},c_{n}) & = & p_{n}\,\,:\,\, P(c_{n})\end{eqnarray*}

In the arity system of the paper, we change the kinds to arities and
the constants and the computation rules are introduced as follows.\begin{eqnarray*}
{\cal T} & : & Zero\\
c_{1} & : & Zero\\
. & . & .\\
c_{n} & : & Zero\\
{\cal E_{{\cal T}}} & : & ((Zero,Zero),\\
 &  & (Zero,(Zero,...(Zero,\\
 &  & (Zero,Zero)...)\end{eqnarray*}
and the following reduction rules\begin{eqnarray*}
{\cal E_{{\cal T}}}(P,p_{1},...,p_{n},c_{1}) & \longrightarrow & p_{1}\,\,:\,\, Zero\\
......\\
{\cal E_{{\cal T}}}(P,p_{1},...,p_{n},c_{n}) & \longrightarrow & p_{n}\,\,:\,\, Zero\end{eqnarray*}
Now, let's consider a concrete example, boolean type. Its representation
in type systems and in the arity system are the following.\begin{eqnarray*}
Bool & : & Type\\
true & : & Bool\\
false & : & Bool\\
{\cal E}_{Bool} & : & (P:(Bool)Type)\\
 &  & (p_{1}:P(true))(p_{2}:P(false))\\
 &  & (z:Bool)P(z)\end{eqnarray*}
\begin{eqnarray*}
{\cal E}_{Bool}(P,p_{1},p_{2},true) & = & p_{1}\,\,:\,\, P(true)\\
{\cal E}_{Bool}(P,p_{1},p_{2},false) & = & p_{2}\,\,:\,\, P(false)\end{eqnarray*}
\begin{eqnarray*}
Bool & : & Zero\\
true & : & Zero\\
false & : & Zero\\
{\cal E}_{Bool} & : & ((Zero,Zero),\\
 &  & (Zero,(Zero,\\
 &  & (Zero,Zero))))\end{eqnarray*}
\begin{eqnarray*}
{\cal E}_{Bool}(P,p_{1},p_{2},true) & \longrightarrow_{b_{1}} & p_{1}\,\,:\,\, Zero\\
{\cal E}_{Bool}(P,p_{1},p_{2},false) & \longrightarrow_{b_{2}} & p_{2}\,\,:\,\, Zero\end{eqnarray*}

\subsection{Universe operator}

We consider some simple case, for example,\begin{eqnarray*}
U & : & Type\\
Bool & : & Type\\
bool & : & U\\
uo & : & (U)Type\end{eqnarray*}
\begin{eqnarray*}
uo(bool) & = & Bool\end{eqnarray*}
\begin{eqnarray*}
U & : & Zero\\
Bool & : & Zero\\
bool & : & Zero\\
uo & : & (Zero,Zero)\end{eqnarray*}
\[
uo(bool)\longrightarrow_{u}Bool\,\,:\,\, Zero\]

\subsection{Strong normalisation w.r.t. $\beta\eta\pi_{1}$-reduction}

We have proved strong normalisation w.r.t. $\beta\eta$-reduction
in Section \ref{sec:Normalisation-proof}. Now, we add the reduction
rule $\pi_{1}$ and prove strong normalisation w.r.t. $\beta\eta\pi_{1}$-reduction.
As mentioned before, the strategy is to prove strong normalisation
one reduction rule after another. So after proving it w.r.t. $\beta\eta\pi_{1}$-reduction,
we can add another rule (\emph{eg, $\pi_{2}$}-reduction), and so
on. In this section, we demonstrate the proof techniques through the
proof w.r.t. $\beta\eta\pi_{1}$-reduction. For other reduction rules
such as $\pi_{2}$, $b_{1}$, $b_{2}$ and $u$, the proof methods
are the same.

\begin{theorem}
If $M$ doesn't have a correct arity under a context $A$ without
the $\pi_{1}$-reduction then $M$ still doesn't have a correct arity
under the context $A$ with the $\pi_{1}$-reduction.
\end{theorem}
\begin{proof}
The arities of the left hand side and the right hand side of the reduction
rule $\pi_{1}$ are the same, and there is no reduction for arities.
So, $\pi_{1}$-reduction becomes irrelevant whether $M$ has a correct
arity.
\end{proof}
\begin{remark}
As mentioned before, in dependently typed systems, a term that is
not well-typed can become a well-typed term after adding new reduction
rules. For instance, under a context $f:(x:B(a))C$ and $y:B(\pi_{1}(pair(a,b)))$,
the term $f(y)$ is not well-typed (some details are omitted here).
However, if we add the $\pi_{1}$-reduction rule, then it becomes
a well-typed term. This example shows that, after adding new reduction
rules, well-typed terms may increase. This is one of the difficulties
to prove the statement that {}``well-typed terms are strongly normalising''.
\end{remark}
Now, in order to prove strong normalisation, we prove some lemmas
first.

\begin{lemma}
\textbf{\label{lem:subst_eta}(Substitution for $\eta$)} If $M_{1}\longrightarrow_{\eta}M_{2}$
then $[N/x]M_{1}\longrightarrow_{\eta}[N/x]M_{2}$. And if $N_{1}\longrightarrow_{\eta}N_{2}$
then $[N_{1}/x]M\twoheadrightarrow_{\eta}[N_{2}/x]M$.
\end{lemma}
\begin{proof}
For the first part, we proceed the proof by induction on $M_{1}$,
and for the second part, by induction on $M$. In the case that $M$
is a variable, we consider two sub-cases: $M\equiv x$ and $M\not\equiv x$.
\end{proof}
\begin{lemma}
\label{lem:free_beta}If $M_{1}\longrightarrow_{\beta}M_{2}$ and
$x\not\in FV(M_{1})$ then $x\not\in FV(M_{2})$.
\end{lemma}
\begin{proof}
By induction on $M_{1}$.
\end{proof}
\begin{lemma}
\label{lem:eta_case}If $M_{1}\longrightarrow_{\eta}\lambda x:K_{2}.M_{2}$
then there are three and only three possibilities as the following:
\begin{itemize}
\item $M_{1}\equiv\lambda y:K_{1}.(\lambda x:K_{2}.M_{2})y$ for some $y$
and $K_{1}$, and $y\not\in FV(\lambda x:K_{2}.M_{2})$.
\item $M_{1}\equiv\lambda x:K_{2}.N$ for some $N$ and $N\longrightarrow_{\eta}M_{2}$.
\item $M_{1}\equiv\lambda x:K_{1}.M_{2}$ for some $K_{1}$ and $K_{1}\longrightarrow_{\eta}K_{2}$. 
\end{itemize}
\end{lemma}
\begin{proof}
By the understanding of one-step reduction.
\end{proof}
\begin{lemma}
\textbf{\label{lem:Commu_eta}(Commutation for $\eta\beta$)} If $M_{1}\longrightarrow_{\eta}M_{2}$
and $M_{2}\longrightarrow_{\beta}M_{3}$ then there exists a $M_{2}'$
such that $M_{1}\twoheadrightarrow_{\beta}^{+}M_{2}'$ and $M_{2}'\twoheadrightarrow_{\eta}M_{3}$.
\end{lemma}
\begin{proof}
By induction on $M_{1}$ and Lemma \ref{lem:subst_eta}, \ref{lem:free_beta}
and \ref{lem:eta_case}.
\end{proof}
\begin{lemma}
\label{lem:subst_pi1}\textbf{(Substitution for $\pi_{1}$)} If $M_{1}\longrightarrow_{\pi_{1}}M_{2}$
then $[N/x]M_{1}\longrightarrow_{\pi_{1}}[N/x]M_{2}$. And if $N_{1}\longrightarrow_{\pi_{1}}N_{2}$
then $[N_{1}/x]M\twoheadrightarrow_{\pi_{1}}[N_{2}/x]M$. 
\end{lemma}
\begin{proof}
Similar to the proof of Lemma \ref{lem:subst_eta}.
\end{proof}
\begin{lemma}
\label{lem:pi1_case}If $M_{1}\longrightarrow_{\pi_{1}}\lambda x:K_{2}.M_{2}$
then there are two and only two possibilities as the following:
\begin{itemize}
\item $M_{1}\equiv\lambda x:K_{2}.N$ for some $N$ and $N\longrightarrow_{\pi_{1}}M_{2}$.
\item $M_{1}\equiv\lambda x:K_{1}.M_{2}$ for some $K_{1}$ and $K_{1}\longrightarrow_{\pi_{1}}K_{2}$. 
\end{itemize}
\end{lemma}
\begin{proof}
By the understanding of one-step reduction and the arity of $M_{1}$
is not $Zero$.
\end{proof}
\begin{lemma}
\textbf{\label{lem:Commu_pi1}(Commutation for $\pi_{1}\beta$)} If
$M_{1}\longrightarrow_{\pi_{1}}M_{2}$ and $M_{2}\longrightarrow_{\beta}M_{3}$
then there exists a $M_{2}'$ such that $M_{1}\longrightarrow_{\beta}M_{2}'$
and $M_{2}'\twoheadrightarrow_{\pi_{1}}M_{3}$.
\end{lemma}
\begin{proof}
By induction on $M_{1}$ and Lemma \ref{lem:subst_pi1} and \ref{lem:pi1_case}.
\end{proof}
\begin{theorem}
If $A\vdash M:a$, then $M$ is strongly normalising w.r.t. $\beta\eta\pi_{1}$-reduction.
\end{theorem}
\begin{proof}
We proceed the proof by contradiction, and by Theorem \ref{thm:SN}
and Lemma \ref{lem:Commu_eta} and \ref{lem:Commu_pi1}. 

Suppose there is an infinite reduction sequence for $M$ and it is
called $S$. By Theorem \ref{thm:SN}, $M$ is strongly normalising
w.r.t. $\beta\eta$-reduction. So, $S$ must contain infinite times
of $\pi_{1}$-reduction. Every time when $\eta$-reduction or $\pi_{1}$-reduction
rule is applied, terms become smaller. So, $M$ is strongly normalising
w.r.t. $\eta\pi_{1}$-reduction. And hence $S$ must also contain
infinite times of $\beta$-reduction. In fact, $S$ must be like the
following,\[
M\twoheadrightarrow_{\eta\pi_{1}}^{+}M_{1}\twoheadrightarrow_{\beta}^{+}M_{2}\twoheadrightarrow_{\eta\pi_{1}}^{+}M_{3}\twoheadrightarrow_{\beta}^{+}M_{4}\twoheadrightarrow_{\eta\pi_{1}}^{+}...\]
or\[
M\twoheadrightarrow_{\beta}^{+}M_{1}\twoheadrightarrow_{\eta\pi_{1}}^{+}M_{2}\twoheadrightarrow_{\beta}^{+}M_{3}\twoheadrightarrow_{\eta\pi_{1}}^{+}M_{4}\twoheadrightarrow_{\beta}^{+}...\]
where $\twoheadrightarrow_{\beta}^{+}$ means one or more but finite
reduction steps of $\beta$, and similarly, $\twoheadrightarrow_{\eta\pi_{1}}^{+}$
means one or more but finite reduction steps of $\eta$ or $\pi_{1}$.

Now, by Lemma \ref{lem:Commu_eta} and Lemma \ref{lem:Commu_pi1},
for the infinite sequence $S$, we can always move the $\beta$-reduction
steps forward and build an infinite sequence of $\beta$-reduction.
This is a contradiction to that $M$ is strongly normalising w.r.t.
$\beta$-reduction.\qed
\end{proof}

\section{\label{sec:Conclusions-and-future}Conclusions and future work}

Strong normalisation for MLF has been proved in the paper, but we
did not follow the traditional understanding, that is, well-typed
terms are strongly normalising. Instead, a weaker condition has been
proposed, which says terms with correct arities are strongly normalising.
The author hopes this new understanding will inspire us to think the
question {}``why is a term strongly normalising?'' again, and to
simplify the proofs for dependently typed systems.

Another important technique employed in the paper is that, in order
to prove what we want, we prove a more general and stronger property.
In the paper, the definition of terms and kinds is extended and a
new reduction rule $\beta_{2}$ is introduced. And we proved strong
normalisation w.r.t. $\beta\eta\beta_{2}$-reduction instead of w.r.t.
$\beta\eta$-reduction only. This generalisation is quite different
from the traditional idea of generalising induction hypothesis.

We only studied the computation rules for some inductive data types
and these rules have commutation property. However, some computation
rules do not have such property, for instance, the computation rule
for the type of function space. How to prove strong normalisation
for such rules needs further study. The question of how to develop
weaker conditions to simplify the normalisation proofs for other type
systems is also worth being taken into our consideration.

\subsubsection*{Acknowledgements}

Thanks to Zhaohui Luo, Sergei Soloviev, James McKinna and Healfdene
Goguen for discussions on the issue of strong normalisation, and for
reading the earlier version of the paper, and for their helpful comments
and suggestions.

\bibliographystyle{alpha}
\bibliography{Yongbib}

\section*{Appendix}

\textbf{Terms and Kinds in MLF}

\begin{itemize}
\item Terms

\begin{enumerate}
\item a variable is a term,
\item $\lambda x:K.M$ is a term if $x$ is a variable, $K$ is a kind and
$M$ is a term,
\item $MN$ is a term if $M$ and $N$ are terms.
\end{enumerate}
\item Kinds

\begin{enumerate}
\item $Type$ is a kind,
\item $El(M)$ is a kind if $M$ is a term,
\item $(x:K_{1})K_{2}$ is a kind if $K_{1}$ and $K_{2}$ are kinds.
\end{enumerate}
\end{itemize}
\textbf{Reduction rules in MLF}

\[
(\lambda x:K.M)N\longrightarrow_{\beta}[N/x]M\]
\[
\lambda x:K.Mx\longrightarrow_{\eta}M\,\,\,\,\,\,\,\,\,\, x\not\in FV(M)\]
\textbf{Inference rules for MLF}

\noindent Contexts and assumptions

\[
\frac{}{<>\:\: valid}\:\:\:\frac{\Gamma\vdash K\:\: kind\:\:\: x\notin FV(\Gamma)}{\Gamma,x:K\:\: valid}\:\:\:\frac{\Gamma,x:K,\Gamma'\:\: valid}{\Gamma,x:K,\Gamma'\vdash x:K}\]
Equality rules\[
\frac{\Gamma\vdash K\:\: kind}{\Gamma\vdash K=K}\:\:\:\frac{\Gamma\vdash K=K'}{\Gamma\vdash K'=K}\:\:\:\frac{\Gamma\vdash K=K'\:\:\Gamma\vdash K'=K''}{\Gamma\vdash K=K''}\]
\[
\frac{\Gamma\vdash k:K}{\Gamma\vdash k=k:K}\:\:\:\frac{\Gamma\vdash k=k':K}{\Gamma\vdash k'=k:K}\:\:\:\frac{\Gamma\vdash k=k':K\:\:\Gamma\vdash k'=k'':K}{\Gamma\vdash k=k'':K}\]
\[
\frac{\Gamma\vdash k:K\:\:\Gamma\vdash K=K'}{\Gamma\vdash k:K'}\:\:\:\frac{\Gamma\vdash k=k':K\:\:\Gamma\vdash K=K'}{\Gamma\vdash k=k':K'}\]
Substitution rules\[
\frac{\Gamma,x:K,\Gamma'\:\: valid\:\:\Gamma\vdash k:K}{\Gamma,[k/x]\Gamma'\:\: valid}\]
\[
\frac{\Gamma,x:K,\Gamma'\vdash K'\:\: kind\:\:\Gamma\vdash k:K}{\Gamma,[k/x]\Gamma'\vdash[k/x]K'\:\: kind}\:\:\:\frac{\Gamma,x:K,\Gamma\vdash K'\:\: kind\:\:\Gamma\vdash k=k':K}{\Gamma,[k/x]\Gamma'\vdash[k/x]K'=[k'/x]K'}\]
\[
\frac{\Gamma,x:K,\Gamma'\vdash k':K'\:\:\Gamma\vdash k:K}{\Gamma,[k/x]\Gamma'\vdash[k/x]k':[k/x]K'}\:\:\:\frac{\Gamma,x:K,\Gamma'\vdash k':K'\:\:\Gamma\vdash k_{1}=k_{2}:K}{\Gamma,[k_{1}/x]\Gamma'\vdash[k_{1}/x]k'=[k_{2}/x]:[k_{1}/x]K'}\]
\[
\frac{\Gamma,x:K,\Gamma'\vdash K'=K''\:\:\Gamma\vdash k:K}{\Gamma,[k/x]\Gamma'\vdash[k/x]K'=[k/x]K''}\:\:\:\frac{\Gamma,x:K,\Gamma'\vdash k'=k'':K'\:\:\Gamma\vdash k:K}{\Gamma,[k/x]\Gamma'\vdash[k/x]k'=[k/x]k'':[k/x]K'}\]
The kind type\[
\frac{\Gamma\:\: valid}{\Gamma\vdash Type\:\: kind}\:\:\:\frac{\Gamma\vdash A:Type}{\Gamma\vdash El(A)\:\: kind}\:\:\:\frac{\Gamma\vdash A=B:Type}{\Gamma\vdash El(A)=El(B)}\]
Dependent product kinds\[
\frac{\Gamma\vdash K\:\: kind\:\:\Gamma,x:K\vdash K'\:\: kind}{\Gamma\vdash(x:K)K'\:\: kind}\:\:\:\frac{\Gamma\vdash K_{1}=K_{2}\:\:\Gamma,x:K_{1}\vdash K_{1}'=K_{2}'}{\Gamma\vdash(x:K_{1})K_{1}'=(x:K_{2})K_{2}'}\]
\[
\frac{\Gamma,x:K\vdash k:K'}{\Gamma\vdash\lambda x:K.k:(x:K)K'}\:\:\:\:\:(\xi)\:\frac{\Gamma\vdash K_{1}=K_{2}\:\:\:\Gamma,x:K_{1}\vdash k_{1}=k_{2}:K}{\Gamma\vdash\lambda x:K_{1}.k_{1}=\lambda x:K_{2}.k_{2}:(x:K_{1})K}\]
\[
\frac{\Gamma\vdash f:(x:K)K'\:\:\Gamma\vdash k:K}{\Gamma\vdash f(k):[k/x]K'}\:\:\:\frac{\Gamma\vdash f=f':(x:K)K'\:\:\Gamma\vdash k_{1}=k_{2}:K}{\Gamma\vdash f(k_{1})=f'(k_{2}):[k_{1}/x]K'}\]
\[
(\beta)\:\frac{\Gamma,x:K\vdash k':K'\:\:\Gamma\vdash k:K}{\Gamma\vdash(\lambda x:K.k')(k)=[k/x]k':[k/x]K'}\:\:\:(\eta)\:\frac{\Gamma\vdash f:(x:K)K'\:\: x\notin FV(f)}{\Gamma\vdash\lambda x:K.f(x)=f:(x:K)K'}\]

\end{document}